\documentclass[aps,prl,twocolumn,groupedaddress,balancelastpage]{revtex4}

\usepackage{graphicx}
\usepackage[letterpaper,total={7in,9.5in},top=0.75in,left=0.75in]{geometry}
\begin{document}

%\preprint{}
\onecolumngrid
\noindent Published in Science Dec 13 2002: 2179-2182.\\
Published online November 7, 2002;10.1126/science.1079107\\
(Submitted 4 October, 2002; accepted 31 October, 2002)\\
Address correspondence to jet@phy.duke.edu
\title{Observation of a
 Strongly-Interacting Degenerate Fermi Gas of Atoms}


\author{K. M. O'Hara}
\author{S. L. Hemmer}
\author{M. E. Gehm}
\author{S. R. Granade}
\author{J. E. Thomas}
\affiliation{Physics Department, Duke University, Durham, North
Carolina 27708-0305}

\begin{abstract}
We report on the observation of a highly-degenerate,
strongly-interacting Fermi gas of atoms. Fermionic $^6$Li atoms in
an  optical trap are evaporatively cooled to degeneracy using a
magnetic field to induce strong, resonant interactions. Upon
abruptly releasing the cloud from the trap, the gas is observed to
expand rapidly in the transverse direction while remaining nearly
stationary in the axial. We interpret the expansion dynamics in
terms of collisionless superfluid and collisional hydrodynamics.
For the data taken at the longest evaporation times, we find that
collisional hydrodynamics does not provide a satisfactory
explanation, while superfluidity is plausible.\end{abstract}

%\keywords{}

\maketitle

As the fundamental constituents of matter  are interacting
fermions, the experimental study  of strongly-interacting,
degenerate Fermi gases will impact theories in fields from
particle physics to materials science.  Although the interactions
between fermions are understood  when they are weak (e.g., quantum
electrodynamics), the treatment of very strong interactions
requires the development of new theoretical approaches. To test
these new approaches, there is a need for experimental systems
with widely tunable interaction strengths, densities, and
temperatures. Ultracold atomic Fermi gases have exactly these
properties, and thus enable tests of calculational techniques for
fundamental systems ranging from quarks in nuclear matter to
electrons in high temperature
superconductors~\cite{Bedaque,Wilczek,Randeria,Stoof,Houbiers,Combescot,Heiselberg,Holland,Timmermans,Griffin,Kokkelmans}.
For this reason, a number of groups are developing methods for
creating and exploring ultracold atomic Fermi
gases~\cite{Jin,Truscott,Salomon,Granade,Hadzibabic,Inguscio}. We
report on the study of a strongly-interacting, degenerate Fermi
gas. In contrast to the isotropic expansion previously observed
for a noninteracting degenerate Fermi gas~\cite{Jin}, we observe
anisotropic expansion when the gas is released from an optical
trap.

An exciting feature of strongly-interacting atomic Fermi gases is
the possibility of high-temperature superfluids that are analogs
of very high temperature
superconductors~\cite{Holland,Timmermans,Griffin,Kokkelmans}. Our
experiments produce the conditions predicted for this type of
superfluid transition. Further, the anisotropic expansion  we
observe has been suggested as a signature of the onset of
superfluidity in a Fermi gas~\cite{Stringari}. We interpret the
observed anisotropic expansion in terms of both collisionless
superfluid hydrodynamics~\cite{Stringari} and a new form of
collisional hydrodynamics.

Strong, magnetically-tunable interactions are achieved in our
experiments by employing a Fermi gas comprising a 50-50 mixture of
the two lowest hyperfine states of $^6$Li, i.e., the $|F=1/2,M=\pm
1/2\rangle$ states in the low-magnetic-field basis. This mixture
has a predicted broad Feshbach resonance near an applied magnetic
field of 860 G~\cite{zerocross,Houbiers12}, where  the energy of a
bound $^6$Li-$^6$Li molecular state is tuned into coincidence with
the total energy of the colliding atoms. This enables the
interaction strength to be widely
varied~\cite{zerocross,Houbiers12,Grimm,JinFB}. It  has also been
suggested that interactions between fermions can be modified by
immersion in a Bose gas~\cite{Modugno}. Our experiments are
performed at 910 G, where the zero energy scattering length $a_S$
is estimated to be $\simeq -10^4\,a_0$ ($a_0=0.53\times 10^{-8}$
cm) and the gas has strongly attractive interactions. Resonance
superfluidity has been predicted to occur at this magnetic field
for sufficiently low temperatures~\cite{Kokkelmans}.

In our experiments, $^6$Li atoms are loaded from a magneto-optical
trap into an ultrastable CO$_2$ laser trap~\cite{Granade}. The
trap oscillation frequencies are $\omega_z=2\pi\times (230\pm 20$
Hz) for the axial (z) direction and $\omega_\perp=2\pi\times
(6625\pm 50$ Hz) for the transverse directions. Rate equation
pumping is used to produce the 50-50 spin-mixture: A broad-band
radio frequency (rf) field centered at 7.4 MHz is applied at a
magnetic field of $\simeq 8$ G, nulling the population difference
according to $\Delta n(t)=\Delta n(0)\,\exp (-2Rt)$, where $R$ is
the pumping rate. In  our experiments, $2R=600\,{\rm sec}^{-1}$;
applying the rf field for $t=0.1$ sec produces precise population
balance.

We achieve very low temperatures via rapid forced evaporation in
the CO$_2$ laser trap. In contrast to experiments which employ
magnetic traps to achieve
degeneracy~\cite{Jin,Truscott,Salomon,Hadzibabic,Inguscio}, this
approach has several natural advantages. First, we are able to
evaporate both  spin states to degeneracy at the desired magnetic
field of 910 G. As a result, the sample is never exposed to fields
near 650 G where loss and heating are
observed~\cite{zerocross,Dieckmann2}. Second, the evaporation
process is identical for both spin states, thereby maintaining the
initial spin balance as well as Fermi surface matching. Third, at
this field, the collision cross section is extremely large and
unitarity-limited, so that runaway evaporation is
expected~\cite{Scaling}.

Forced evaporation is achieved by lowering the power of the
trapping laser, while maintaining the beam profile and angular
alignment. The trap depth $U$ is reduced for 3.5 s according to
the trajectory $U(t)=U_0\,(1+t/\tau)^{-1.45}+U_B$~\cite{Scaling},
where $U_B$ is a small offset. The value of $\tau$ is taken to be
0.1 s, large compared to the time constant estimated for achieving
degeneracy at 910 G. With this choice, very high evaporation
efficiency is achieved, yielding extremely low temperatures.

After evaporation, the trap is adiabatically recompressed to full
depth over 0.5 s and then held for 0.5 s to ensure thermal
equilibrium. While maintaining the applied magnetic field of 910
G, the gas is released from the trap and imaged at various times
to observe the anisotropy. The CO$_2$ laser power is extinguished
in less than 1 $\mu$s with a rejection ratio of $2\times
10^{-5}$~\cite{Granade}.

A CCD camera images the  gas from a direction perpendicular to the
axial direction (z)  of the trap and parallel to the  applied
magnetic field direction (y). The small repulsive potential
induced by the high-field magnet is along the camera observation
axis and does not affect the images. The remaining attractive
potential has cylindrical symmetry and corresponds to a harmonic
potential with an oscillation frequency for $^6$Li of 20 Hz at 910
G. Resonant absorption imaging is performed on a cycling
transition at a fixed high magnetic field by using a weak
($I/I_{\rm sat}\simeq 0.05$), 20 $\mu$s probe laser  pulse that is
$\sigma_-$ polarized with respect to the y-axis. Any residual
$\sigma_+$ component is rejected by an analyzer. At 910 G, the
transitions originating from the two occupied spin states are
split by 70 MHz and are well resolved compared to the 3 MHz
half-linewidth, permitting precise determination of the column
density and hence the number of atoms per state. The magnification
is found to be $4.9\pm 0.15$ by moving the axial position of the
trap through 0.5 mm with a micrometer. The net systematic error in
the number measurement is estimated to be
$^{+10\%}_{-6\%}$~\cite{error}. The spatial resolution is
estimated to be $\simeq 4\,\mu$m by quadratically combining the
effective pixel size, 13.0 $\mu$m/4.9, with the aperture limited
spatial resolution of $\simeq 3\,\mu$m.

Figure~1 shows  images of the anisotropic expansion of the
degenerate gas at various times $t$ after release from full trap
depth. The gas rapidly expands in the transverse direction
(Fig.~2A) while remaining nearly stationary in the axial direction
(Fig.~2B) over a time period of 2.0 ms. In contrast to ballistic
expansion, where the column density is $\propto 1/t^2$, the column
density decreases only as $1/t$ for anisotropic expansion.
Consequently, the signals are quite large even for long expansion
times.

\begin{figure}
\includegraphics[width=50mm]{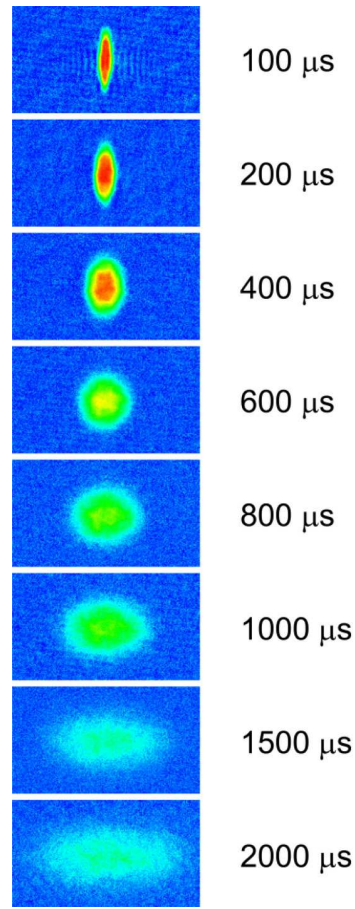}
\caption{False-color absorption images of a strongly interacting,
degenerate Fermi gas as a function of time $t$ after release from
full trap depth for $t=0.1-2.0$ ms, top to bottom.  The axial
width of the gas remains nearly stationary as the transverse width
expands rapidly.} \label{fig:1}
\end{figure}

\begin{figure}
\includegraphics[width=75mm]{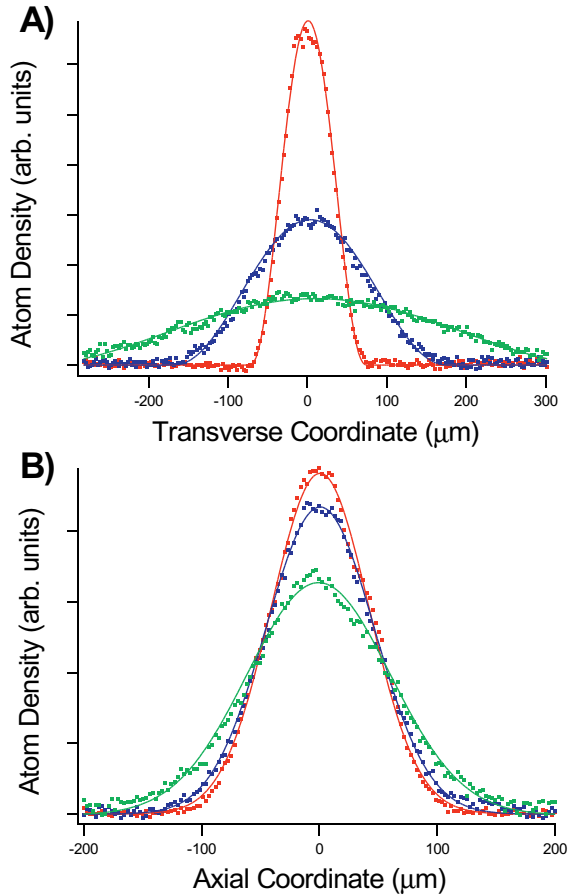}
\caption{One-dimensional spatial distributions in the transverse
({\bf A}) and axial ({\bf B}) directions (red, 0.4 ms; blue, 1.0
ms; green, 2.0 ms). The transverse distributions are shown fit
with zero-temperature Thomas-Fermi distributions, while the axial
are shown fit with gaussian distributions.} \label{fig:2}
\end{figure}

One possible explanation of the observed anisotropy is provided by
a recent theory of collisionless superfluid
hydrodynamics~\cite{Stringari}. After release from the trap, the
gas expands hydrodynamically due to the force from an effective
potential $U_{\rm eff}=\epsilon_F+U_{\rm MF}$, where
$\epsilon_F({\bf x})$ is the local Fermi energy
 and $U_{\rm MF}({\bf x})$ is the mean field contribution. In general, $U_{\rm
MF}\propto a_{\rm eff}\,n$, where $n$ is the spatial density and
$a_{\rm eff}$ is an effective scattering length. However, this
theory is not rigorously applicable to our experiment, as it was
derived for the dilute limit assuming a momentum-independent
scattering length $a_{\rm eff}=a_S$. This assumption is only valid
when $k_F|a_S|<1$, where $k_F$ is the Fermi wavevector. By
contrast, our experiments are performed in the intermediate
density regime~\cite{Heiselberg}, where $k_F|a_S|\gg 1$, and the
interactions are unitarity-limited. We have therefore attempted to
extend the theory  in the context of a simple model. We make the
assumption that unitarity limits $a_{\rm eff}$ to $\simeq 1/k_F$.
As $n\propto k_F^3$ and $\epsilon_F({\bf x})=\hbar^2 k_F^2/(2M)$,
we obtain $U_{\rm MF}=\beta\, \epsilon_F({\bf x})$, where $\beta$
is a constant. This simple assumption  is further justified by
more detailed calculations~\cite{Heiselberg} which show that
$\beta$ is an important universal many-body parameter. With this
assumption, $U_{\rm eff}({\bf x})=(1+\beta )\, \epsilon_F({\bf
x})$. As $\epsilon_F\propto n^{2/3}$,  it then follows that
$U_{\rm eff}\propto n^{2/3}$.

For  release from a harmonic trap and $U_{\rm eff}\propto
n^\gamma$, the hydrodynamic equations admit an exact
solution~\cite{Stringari,Kagan},
\begin{equation}
n({\bf x},t)=\frac{n_0(x/b_x,y/b_y,z/b_z)}{b_xb_yb_z}.
\label{eq:scaling}
\end{equation}
Here, $n_0({\bf x})$ is the initial spatial distribution in the
trap, and $b_i(t)$ are time dependent scaling parameters, which
satisfy simple coupled differential equations with the initial
conditions, $b_i(0)=1$, $\dot{b}_i(0)=0$, where $i=x,y,z$. As the
shape of the initial distribution (even with the mean field
included) is determined by the trap potential,  $n_0$ is a
function only of $r'$ where $\bar{\omega}^2
r'^2=\omega_\perp^2(x^2+y^2) +\omega_z^2z^2$ and
$\bar{\omega}=(\omega_\perp^2\omega_z)^{1/3}$. Hence, the initial
radii of the density distribution $n_0$ are in the proportion
$\sigma_x(0)/\sigma_z(0)=\lambda\equiv\omega_z/\omega_\perp$. For
our trap, $\lambda=0.035$. Then, during hydrodynamic expansion,
the  radii of the density distribution evolve according to
\begin{equation}
\sigma_x(t)=\sigma_x(0)\,b_x(t),\quad\sigma_z(t)=\sigma_z(0)\,b_z(t).
 \label{eq:ratio}
\end{equation}
We determine $b_x(t)=b_y(t)$ and $b_z(t)$ from their evolution
equations~\cite{Stringari}. For $\gamma =2/3$,
$\ddot{b}_x=\omega_\perp^2\,b_x^{-7/3}b_z^{-2/3}$ and
$\ddot{b}_z=\omega_z^2\,b_x^{-4/3}b_z^{-5/3}$.

>From the expansion data, the  widths $\sigma_x(t)$ and
$\sigma_z(t)$ are determined by fitting  one-dimensional
distributions~\cite{projected} with normalized, zero-temperature
Thomas-Fermi (T-F) distributions,
$n(x)/N=(16/(5\pi\sigma_x))(1-x^2/\sigma_x^2)^{5/2}$. As shown in
Fig.~2A, the zero-temperature T-F fits to the transverse spatial
profiles are quite good. This shape is not unreasonable despite a
potentially large mean field interaction. As noted above $U_{\rm
MF}({\bf x})\propto\epsilon_F({\bf x})$. Hence, the mean field
simply rescales the Fermi energy in the equation of
state~\cite{Stringari}. In this case, it is easy to show that the
initial shape of the cloud is expected to be that of a
Thomas-Fermi distribution. This shape is then maintained by the
hydrodyamic scaling of Eq.~\ref{eq:ratio}.

Figure~3A shows the measured values of $\sigma_x(t)$ and
$\sigma_z(t)$ as a function of time $t$ after release. To compare
these results with the predictions of Eq.~\ref{eq:ratio}, we take
the initial dimensions of the cloud, $\sigma_x(0)$ and
$\sigma_z(0)$, to be the zero-temperature Fermi radii. For our
measured number $N=7.5^{+0.8}_{-0.5}\times 10^4$ atoms per state,
and $\bar{\omega}=2\pi\times (2160\pm 65$ Hz), the Fermi
temperature is
$T_F=\hbar\bar{\omega}(6N)^{1/3}/k_B=7.9^{+0.3}_{-0.2}\,\mu$K at
full trap depth. One then obtains
$\sigma_x(0)=\sqrt{2k_BT_F/M\omega_x^2}=3.6\pm 0.1\,\mu$m in the
transverse direction, and $\sigma_z(0)=103\pm 3\,\mu$m in the
axial. For these initial dimensions, we obtain very good agreement
with our measurements using no free parameters, as shown by the
solid curves in Fig.~3A.

Figure~3B shows the measured aspect ratios
$\sigma_x(t)/\sigma_z(t)$ and the theoretical predictions  based
on hydrodynamic, ballistic, and attractive ($\beta =-0.4$) or
repulsive ($\beta =0.4$) collisionless mean field
scaling~\cite{Stringari}. The observed expansion appears to be
nearly hydrodynamic. For comparison, we also show the measured
aspect ratios obtained for release at 530 G, where the scattering
length has been measured to be nearly zero~\cite{zerocross,Grimm}.
In this case, there is excellent agreement with the ballistic
expansion expected for a noninteracting gas. This directly
confirms that the observed anisotropy is a consequence of
interactions.

\begin{figure}
\includegraphics[width=75mm]{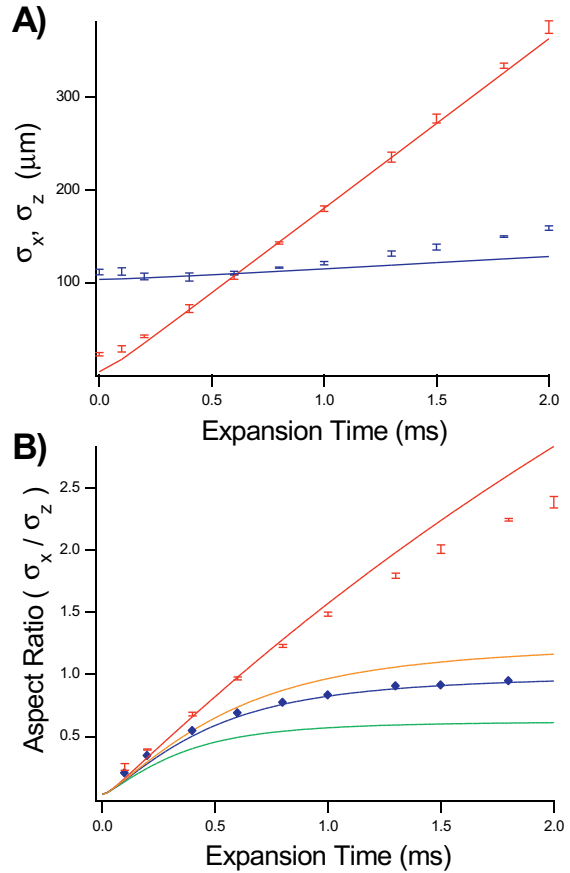}
\caption{({\bf A}) Transverse (red) and axial (blue) widths as
functions of time after release. The solid curves are theoretical
predictions based on hydrodynamic scaling with no free parameters.
({\bf B}) Aspect ratio of the cloud as a function of time after
release. The dots indicate experimental data and the solid curves
show theoretical predictions with no adjustable parameters (red,
hydrodynamic; blue, ballistic; green, attractive mean field;
orange, repulsive mean field). } \label{fig:3}
\end{figure}

A primary assumption of our simple model is that the system is
unitarity limited, i.e., $|k_F\,a_S|\gg 1$.  In this limit the
fluid properties are predicted to be independent of $k_Fa_S$, for
both a collisionless superfluid~\cite{Heiselberg} and for
collisional hydrodynamics as shown below (see
Eq.~\ref{eq:meanfq}). To test this assumption, we have released
the gas at a trap depth $U$ which is 1/100 of the full trap depth.
As the trap frequencies, and hence the Fermi energy, scale as
$\sqrt{U}$, $k_F$ is reduced by $(1/100)^{1/4}\simeq 1/3$. The
time scale for the expansion is increased by a factor of 10 as
expected, and we continue to observe strongly anisotropic
expansion. This suggests that the system is unitarity-limited at
full trap depth, consistent with the estimated value of
$k_F|a_S|=7.4$ for our experimental conditions. Hence, the gas
appears to be strongly interacting.

In the unitarity limit, where $U_{\rm MF}=\beta\,\epsilon_F$, we
can obtain the first estimate of $\beta$ from measurements of the
transverse release energy. From the zero-temperature T-F fits to
the 910 G data, we obtain an average transverse release energy of
$\epsilon_x=\langle
Mv_x^2/2\rangle=M(\sigma_x/4t)^2=1.44\pm0.02\,\mu$K per particle.
To derive an expression for the release energy as a function of
$\beta$, we assume that the initial density distribution $n_0({\bf
x})$ is determined by the  equation of state for a trapped, normal
fluid~\cite{Stringari}: $\epsilon_F({\bf x})+U_{\rm MF}({\bf x})+
U_{\rm trap}({\bf x})=\mu$, where $\mu$ is the chemical potential.
In this case, the release energy per particle is
    \begin{equation}
    \epsilon_r=\frac{3}{8}\,k_BT_F\,\sqrt{1+\beta}.
    \end{equation}
As shown above, $T_F=7.9^{+0.3}_{-0.2}\,\mu$K at full trap depth.
To obtain an estimate of $\beta$ from the experiments, we assume
$\epsilon_r/2$ is released in each transverse direction, with
negligible energy deposited axially. Calculating $\beta$ using the
appropriate $T_F$ and $\epsilon_x$ for each trial in the range
$t=0.4-0.8$ ms, we find $\beta =-0.10\pm 0.07$. Note that the
uncertainty is the quadratic combination of the statistical and
systematic uncertainties in the  measurements, but does not
reflect any systematic effects arising from our model. We find
that the sign is in agreement with recent predictions, but the
magnitude is a factor of ten smaller than
expected~\cite{Heiselberg}. It is remarkable that the release
energy and the initial cloud dimensions are well-described
assuming a zero-temperature, noninteracting Fermi gas, while the
highly anisotropic expansion results from strong interactions.

The preceding determination of $\beta$ assumed a cloud  at zero
temperature. To investigate the validity of this approximation, we
fit normalized, finite-temperature T-F distributions to the
transverse data at short times, 0.2 and 0.4 ms, where the signal
to background ratio in the thermal wings is high. As noted above,
it is not unreasonable to expect a T-F distribution. In this case,
it is possible to perform a two parameter fit with $\sigma_x$ and
$T/T_F$ as the free parameters, where $T/T_F$ is the ratio of the
temperature to the Fermi temperature.  The fits (not shown) to the
measured transverse spatial distributions yield $0.08\leq
T/T_F\leq 0.18$, while $\sigma_x$ is essentially unchanged from
the zero-temperature results. Hence, it appears that $T/T_F$ is
quite small and the zero-temperature approximation is reasonable.

In our experiments, the peak Fermi density is calculated to be
$n_F=4.7\times 10^{13}/{\rm cm}^{3}$ per state~\cite{Butts}. At
this density, it is possible that  the anisotropic expansion
arises from collisional hydrodynamics~\cite{Kagan}. In contrast to
the case usually considered, for the large scattering lengths in
this system, the collision cross section  is unitarity-limited to
a value $\sigma_F\simeq 4\pi/k_F^2$ in the degenerate regime. The
gas is collisionally hydrodynamic when the collision parameter
$\phi=\gamma/\omega_\perp\gg 1$, where $\gamma$ is the elastic
collision rate and $1/\omega_\perp$ is the relevant time scale for
the expansion. Using the s-wave Boltzmann equation~\cite{Walraven}
for a constant cross section $\sigma$ and including Pauli
blocking~\cite{JinPB}, we find that $\gamma =\gamma_0\,F_P(T/TF)$,
where $\gamma_0=N M\sigma\bar{\omega}^3/(2\pi^2\,k_BT_F)$, $N$ is
number of atoms in one spin state, and $F_P$ describes the
temperature dependence. Here, $F_P\rightarrow T_F/T$ for $T\gg
T_F$, and $F_P\simeq 15\,(T/T_F)^2$ for $T<0.2\, T_F$, where Pauli
blocking occurs. The maximum value of $F_P=1.3$ occurs at
$T/T_F=0.5$. In the degenerate regime (i.e., $T<0.5\,T_F$), where
$\sigma \simeq \sigma_F$, we obtain
\begin{equation}
\phi=\frac{(6\lambda N)^{1/3}}{6\pi}F_P(T/T_F). \label{eq:meanfq}
\end{equation}
For our experimental conditions, $\phi\simeq 1.3\,F_P(T/T_F)$,
which is independent of the trap depth as long as $k_F|a_S|\gg 1$.
As $F_P$ is at most of order unity, strongly hydrodynamic behavior
arising from collisions seems unlikely. Including the temperature
dependence, $\phi$ ranges from 0.8 down to 0.2 where the system is
nearly collisionless. Hence, collisional hydrodynamics does not
provide a satisfactory explanation of the observed anisotropic
expansion, while superfluid hydrodynamics is plausible.

Given this  possibility, we have performed an initial
investigation of the transition between ballistic and hydrodynamic
expansion. We measure the aspect ratio for an expansion time of
0.6 ms as a function of the evaporation time. For short
evaporation times $< 0.13$ s, where $T/T_F>3.5$, the measured
aspect ratio is consistent with that expected for ballistic
expansion. For any evaporation time $>\,1.5$ s, the aspect ratio
is consistent with hydrodynamic expansion. We observe a very
smooth transition between these two extremes. In the intermediate
regime, at temperatures below $T/T_F=3.5$,  the expansion lies
between hydrodynamic and ballistic. At $T/T_F=3.5$, where the
evaporation time is short and the number is large, an estimate of
the classical collision rate with a unitarity-limited cross
section shows that the onset of collisional behavior is not
surprising. In the intermediate region, there is no theory of
expansion to describe the spatial anisotropy of the energy
release. Hence, any attempt to determine the temperature is highly
model-dependent and cannot be trusted. To further complicate the
analysis, varying the evaporation time changes the trap population
in addition to the temperature. Finally, if high-temperature
resonance superfluidity does exist, the transition temperature is
predicted to be in the range 0.25-0.5 $T_F$, where  Pauli blocking
is not very effective. Hence, one would not expect to observe a
collisionless region immediately prior to the onset of superfluid
hydrodynamics, unless the transition occurs at very low
temperature, in contrast to predictions.

There are a number of noticeable  discrepancies between the
hydrodynamic theory and the data.  The deviations at 0 and 0.1 ms
can be explained by possible index of refraction effects as well
as spatial resolution limits. These issues are not significant for
longer expansion times where the density is reduced and the cloud
size is well beyond the resolution limit of our imaging system.
However, close examination of the long time deviations reveals
that there may be a two-component structure in the gas.  In the
axial direction, hydrodynamic expansion is very slow, and a second
component expanding according to ballistic or collisionless mean
field scaling (Fig.~3B) easily overtakes the hydrodynamic
component. A two-component structure may also explain why  the
axial spatial distributions (Fig.~2B), are better fit by gaussian
distributions than by zero-temperature T-F distributions. By
contrast, in the transverse direction, the hydrodynamic expansion
is the fastest, masking any two-component structure after a short
time.

\begin{acknowledgments}
We are indebted to H. Heiselberg and S. Stringari for
stimulating correspondence as well as for independently providing
us with the release energy formula of Eq. 3. We also thank
Professor D. Gauthier for a critical reading of the manuscript.
This research is supported by the Chemical Sciences, Geosciences
and Biosciences Division of the Office of Basic Energy Sciences,
Office of Science, U. S. Department of Energy through a program on
the dynamics of ultracold Fermi gases and by the National Science
Foundation through a program on resonance superfluidity. We are
grateful for continuing support from the Physics Division of the
Army Research Office  and the Fundamental Physics in Microgravity
Research program of the National Aeronautics and Space
Administration.
\end{acknowledgments}


\begin{thebibliography}{99}
\bibitem{error}
The systematic uncertainty in the number arises from errors in
detuning,
  saturation parameter, and residual optical pumping (which underestimate the
  true number, as we assume the maximum cross section), as well as errors in
  the magnification.
\bibitem{projected}
Note that the one-dimensional transverse (axial) distributions are
obtained by
  integrating the column density in the axial (transverse) direction.
\end{thebibliography}
\end{document}